\documentclass{article}

\usepackage{arxiv}

\usepackage{amsmath}
\usepackage{algorithmic}
\usepackage{graphicx}
\usepackage{textcomp}
\usepackage{xcolor}
\usepackage{adjustbox}
\usepackage{multirow} 
\usepackage[normalem]{ulem}

\usepackage[utf8]{inputenc} 
\usepackage[T1]{fontenc}    
\usepackage{hyperref}       
\usepackage{url}            
\usepackage{booktabs}       
\usepackage{amsfonts}       
\usepackage{nicefrac}       
\usepackage{microtype}      
\usepackage{cleveref}       
\usepackage{lipsum}         
\usepackage{graphicx}
\usepackage{doi}

\title{An Integrated Failure and Threat Mode and Effect Analysis (FTMEA) Framework with Quantified Cross-Domain Correlation Factors for Automotive Semiconductors }

\usepackage{authblk}

\newbox{\orcid}\sbox{\orcid}{\includegraphics[scale=0.06]{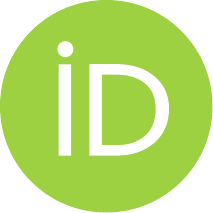}} 
\author[1]{%
	\href{https://orcid.org/0009-0008-9765-636X}{\usebox{\orcid}\hspace{1mm}Antonino Armato\thanks{\texttt{antonino.armato@it.bosch.com}}}%
}
\author[2]{%
	\href{https://orcid.org/0000-0002-3839-1575}{\usebox{\orcid}\hspace{1mm}Marzana Khatun\thanks{\texttt{marzana.khatun@de.bosch.com}}}%
}
\author[2]{%
	{\usebox{\orcid}\hspace{1mm}Sebastian Fischer\thanks{\texttt{sebastian.fischer4@de.bosch.com}}}%
}

\affil[1]{Functional Safety, Robert Bosch GmbH, Milan,Italy}
\affil[2]{Functional Safety Robert Bosch GmbH, Reutlingen, Germany}

\renewcommand{\shorttitle}{\textit{} }

\begin{document}
\maketitle

\begin{abstract}
The automotive industry faces increasing challenges in ensuring both functional safety (FuSa) and cybersecurity for complex semiconductor devices. Traditional Failure Mode and Effects Analysis (FMEA) primarily addresses safety-related failure modes, often overlooking synergistic vulnerabilities and shared consequences with cybersecurity threats. This paper introduces an Integrated Failure and Threat Mode and Effect Analysis  (FTMEA) framework that systematically co-analyzes FuSa and cybersecurity. A cornerstone of this framework is the introduction of rigorously defined Cross-Domain Correlation Factors (CDCFs), which quantify the interdependencies and mutual influences between safety-related failures and cybersecurity threats. These factors are derived from a combination of structured expert knowledge, static structural analysis metrics (e.g., Controllability/Observability), and validated against empirical data from fault/attack injection campaigns, offering a verifiable and reproducible basis for their values. We propose a modified Risk Priority Number (RPN) calculation that systematically integrates these correlation factors, enabling a more accurate and transparent prioritization of risks that span both domains.  A detailed case study involving an automotive Application Specific Integrated Circuit (ASIC) configuration register demonstrates the practical application of the FTMEA. We present explicit mapping tables, quantitative CDCF values, and a comparative analysis against a baseline FMEA/TARA (Threat Analysis and Risk Assessment), illustrating how the integrated approach uncovers previously masked cross-domain risks, improves mitigation strategy effectiveness, and provides a clear quantitative justification for the derived correlation values. This framework offers a unified, traceable,  methodology for risk assessment in critical automotive systems, thereby overcoming the limitations of conventional analyses and promoting optimized, cross-disciplinary development.
\end{abstract}

\section{Introduction}
\subsection{Context and Problem Statement}
The escalating complexity and interconnectedness of modern automotive systems, particularly within semiconductor devices, necessitate a paradigm shift in reliability engineering. Functional Safety (FuSa), governed by standards like ISO 26262 and ISO 21448, ensures freedom from unreasonable risk due to hardware faults or functional insufficiencies. Concurrently, cybersecurity, formalized by ISO/SAE 21434 \cite{b2}, safeguards against malicious attacks and system breaches. Historically, these domains have often been analyzed in silos using distinct methodologies such as FMEA for safety and various threat analyses for security.
However, the increasing overlap between safety-critical functions and cybersecurity vulnerabilities—where a cyberattack can directly compromise safety, or a safety mechanism can inadvertently introduce a security vulnerability—demands a holistic approach. Conventional FMEA, while effective for hardware failure modes, is ill-equipped to systematically capture and quantify these intricate cross-domain dependencies. The absence of a unified framework leads to fragmented analyses, suboptimal risk prioritization, and potentially overlooked vulnerabilities, increasing the likelihood of emergent safety hazards arising from cyber threats.

\subsection{State-of-the-Art and Identified Gaps}
While recent standards and guidelines (e.g., ISO/TS 5083:2025  \cite{b33}, IEC TR 63069 \cite{b7}, SAE J3061 \cite{b5}) acknowledge the interaction between FuSa and cybersecurity, they often lack prescriptive, quantitative methodologies for their integrated analysis. Existing co-analysis methods typically fall into two categories: 
\begin{itemize}
	\item qualitative scenario-based approaches, and
           \item extensions of FMEA (e.g., FMVEA \cite{b14}, FVMEARA \cite{b15}) that introduce security aspects but often struggle with the rigorous quantification of interdependencies and validation of their augmented risk metrics. 
\end{itemize}
A significant gap persists in:
\begin{itemize}
\item  Providing a quantified, traceable, and empirically justifiable mechanism to model the correlations and mutual influences between functional failures and cybersecurity threats.
\item  Developing a mathematical methodology that overcomes ambiguities in defining new risk parameters.
\item  Establishing a systematic validation strategy that includes comparative analysis against established baselines and transparent justification of derived numerical values.
\end{itemize}

\subsection{Main Contributions}
This paper addresses these critical gaps by proposing a novel Integrated Failure and Threat Mode and Effect Analysis (FTMEA) framework that significantly advances the state-of-the-art in several key aspects:
\begin{itemize}
\item Quantified Cross-Domain Correlation Factors (CDCFs): We introduce a novel set of CDCFs that rigorously quantify the mutual influence and shared consequences between functional safety Failure Modes (FMs) and cybersecurity Threat Modes (TMs). These factors are derived through a transparent methodology combining structured expert elicitation, static structural analysis (using metrics like Cone of Influence and Controllability/Observability derived from gate-level netlists), and validated against empirical fault/attack injection campaign data. This provides a verifiable basis for their numerical values, directly addressing previous concerns about ambiguity and lack of rationale.
\item  Mathematically Robust RPN Enhancement with Rescaling: We present a modified Risk Priority Number (RPN) calculation that systematically integrates these CDCFs into the Occurrence (O) and Detection (D) ratings. This enhanced RPN provides a more accurate, reproducible, and justifiable prioritization of risks. Importantly, a systematic rescaling method is applied to map the corrected O and D values back to the discrete 1-10 FMEA ordinal scale \cite{b11}, ensuring practical applicability while overcoming previous ambiguities in scaling, additive structures, and arbitrary truncation.
\item Operationalized Methodology for Integrated Analysis: The framework provides clear, operational definitions and steps for integrating FuSa and cybersecurity considerations throughout the analysis lifecycle, from hazard identification to mitigation strategy evaluation.
\item Empirically Validated Case Study: We demonstrate the FTMEA framework through a detailed case study of an automotive ASIC configuration register. This validation includes explicit mapping tables, quantitative CDCF values with their derivation, a comprehensive comparative analysis against a baseline FMEA/TARA, and a discussion of the insights gained regarding risk prioritization and mitigation effectiveness. This directly addresses the lack of validation and reproducibility concerns.
\item Improved Cross-Disciplinary Collaboration: By providing a common quantitative language and structured methodology, the FTMEA framework inherently optimizes collaboration between safety and cybersecurity teams, leading to more resilient and secure designs. This addresses the need for clear differentiation and validated advancements over existing methods.
\end{itemize}

\subsection{Paper Structure}
The remainder of this paper is organized as follows: Section 2 provides a detailed review of current FuSa and cybersecurity standards and co-analysis methodologies, rigorously identifying their limitations and thus establishing the clear novelty of our approach. Section 3 elaborates on the proposed FTMEA framework, detailing the definition and rigorous quantification of correlation factors, the mathematically justified modified RPN calculation, and the subsequent rescaling procedure. Section 4 presents a comprehensive case study, including explicit data, CDCF derivation, and comparative analysis against baseline methods. Section 5 discusses the implications, limitations, and future work.

\section{State of the Art: Integrated FuSa and Cybersecurity Analysis}

\subsection{Standards and Interactions}
Functional Safety and Cybersecurity aim to mitigate system risks but target distinct hazards. 
Functional Safety, governed by 
standards such as ISO 26262, addresses unintentional risks arising from system vulnerabilities, aiming to ensure the absence of unreasonable risk due to 
hazards caused by malfunctioning behaviors of the Electrical/Electronic (E/E) systems \cite{b1}. 
In addition, ISO 21448 addresses hazardous events caused by functional 
insufficiencies in E/E systems, artificial intelligence-based algorithms, human–machine interfaces, 
and environmental or human factors, aiming to ensure the absence of unreasonable risk from performance 
limitation of the intended functionality or its implementation \cite{b8}.
Similarly, cybersecurity has gained prominence in the automotive industry, 
particularly with the 
emergence of connected and autonomous vehicles, which has led to the 
introduction of standards such as 
ISO/SAE 21434 to protect against external threats \cite{b2}. 
Safety and security often conflict, requiring prioritization strategies. Success depends on aligning 
objectives, ensuring consistency, and addressing goal conflicts highlighted by ISO 26262
\cite{b1}. However, ISO/SAE 21434 \cite{b2} lacks focus on safety interfaces, underscoring the need to 
synchronize safety and security lifecycles early in the design phase.
For example, safety systems focus on predictable safe behavior, while 
security systems should be 
dynamic and adaptive to prevent external breaches. Ensuring that 
safety-critical systems remain safe 
without compromising their FuSa leads to complex design and implementation 
issues.
The challenge is not only to recognize the potential trade-offs,
but to create a coherent development framework that
harmonizes both aspects without increasing system complexity
or development costs.
A need for an integrated
safety and security analysis is also widely acknowledged, by
the standards where the interaction between FuSa and cybersecurity is increasingly recognized by methods that incorporate both domains:
\begin{itemize}
\item IEC TR 63069 \cite{b7}. Security implementations and safety implementation shall be 
compatible. 
Therefore, security countermeasures should effectively prevent or guard against adverse impacts of 
threats to safety-related systems and their implemented safety functions and 
vice versa.
\item SAE J3061 \cite{b5}, ISO/TR	4804 \cite{b6} support the interaction between safety and 
cybersecurity and describe methods and practices for dealing with this overlap.
\item ISO 24089 \cite{b9}	and UN Regulation No. 156 \cite{b23} address safety and security in 
software 
update.
\item ANSI/UL 4600 \cite{b24} address safety evaluation of autonomous vehicles that complement 
other standards with 
safety performance indicator on system-level safety case validity.
\item  UN Regulation No. 155 \cite{b25} concern the approval of vehicles with regards to 
cybersecurity and 
cybersecurity management system including the risk assessment knowledge.
\end{itemize}
While these standards universally acknowledge the critical need to consider both functional safety and cybersecurity, they primarily provide qualitative guidance or establish high-level requirements \cite{b28}. They generally lack prescriptive, quantitative methodologies for systematically assessing the mutual impacts, enabling relationships, and shared consequences between failures and threats. This absence often leads to subjective interpretations in practice, hindering systematic and reproducible integrated risk assessment.

\subsection{Methods and Techniques}
According to \cite{b4}, the relationship between safety and 
cybersecurity has four criteria as :i) Mutual, ii) conditional, iii) independent and iv) Antagonistic 
at high level overview. These criteria has infulence in the safety and cybersecurity analyses. The 
proposed 
FTMEA approach focuses on mutual relationship between them.
There are several methods available, presented in various publications and 
standards, that can help to address the interaction between functional safety and 
cybersecurity. 
Table \ref{tab1} outlines a number of methods and techniques that can be used for 
both safety and security analysis, highlighting that some approaches need to be extended for 
preforming co-analysis effectively, evaluating their limitations regarding quantified and reproducible integrated analysis.
\begin{table}[htb]
	\centering
{\caption{Safety and Security Analysis Methods and Their Quantitative Integration Gap}
\label{tab1}
\begin{tabular}{|c|c|c|c|}
\hline
\textbf{Methods and Techniques} & \multicolumn{3}{c|}{\textbf{Domain:}}\\
	\cline{2-4}
	& \textbf{Safety} & \textbf{Security}& \textbf{Gap} \\
\hline
FMEA:Failure Mode and Effect  & X &{}  &{Not quantify cybersecurity }\\
Analysis \cite{b1,b12,b13} & & & {influence on safety} \\
\hline
FMVEA:Failure Mode,   & X &{X}  &{Qualitative mapping or} \\
Vulnerabilities and Effects &  &{} &{ expert judgment }  \\
Analysis\cite{b14} &  &{} &{for interdependencies}  \\
\hline
FVMEARA:Failure and Vulnerability  & X &{X} & {Qualitative mapping or}  \\
Modes, and Effect Analysis &  &{}  &{expert judgment for}\\
{ and Risk Assessment\cite{b15}}  & X &{X} &{interdependencies}  \\
\hline
ETA:Event Tree Analysis \cite{b16}& X &{X}  &{Not quantifying their relationships } \\
\hline
STPA-SafeSec:System Theoretic Process  & X &{X} &{Requiring significant effort }  \\
Analysis-Safety and Security Analysis \cite{b17} &  &{} &{to quantify the impacts } \\
\hline
SafeSoCPS:Safe System of  & X &{X}  &{Often conceptual or high-level} \\
Cyber-Physical Systems\cite{b18} &  &{}  &{} \\
\hline
AVES:Automated Vehicle Safety and & X &{X} &{It may not provide the precise}  \\
Security Analysis Framework\cite{b19} & & &{quantitative methods } \\
\hline
FTA:Fault Tree Analysis\cite{b1,b16} & X &{}  &{Not quantifying their relationships} \\
\hline
HEAVENS / EVITA: risk-based model\cite{b5} & {} &{X} &{Security-centric scoring systems }  \\
\hline
CVSS:Common Vulnerability  & {} &{X} &{Security-centric scoring systems } \\
Scoring System\cite{b2,b5} & & &{} \\
\hline
\end{tabular}
}
\end{table}

Despite the widespread adoption of various Failure Modes and Effects Analysis methodologies, a significant gap remains in the quantitative co-analysis of FuSa and cybersecurity risks. Existing methods often fall short in providing a rigorously defined, empirically or analytically justified, and quantitatively verifiable mechanism to model the intricate, bidirectional influences between functional failures and cybersecurity threats. This leads to subjective interpretation of "common effects," "mutual influences," and "prioritization," hindering reproducibility and robust decision-making. Our proposed FTMEA framework directly targets this critical gap by introducing quantified correlation factors and a revised RPN calculation, addressing the limitations identified above.

\section{Proposed approach}

\subsection{Co-Analysis Concept including Correlation factors} 
\begin{figure}[htbp]
	\centering
	\includegraphics[scale=0.50]{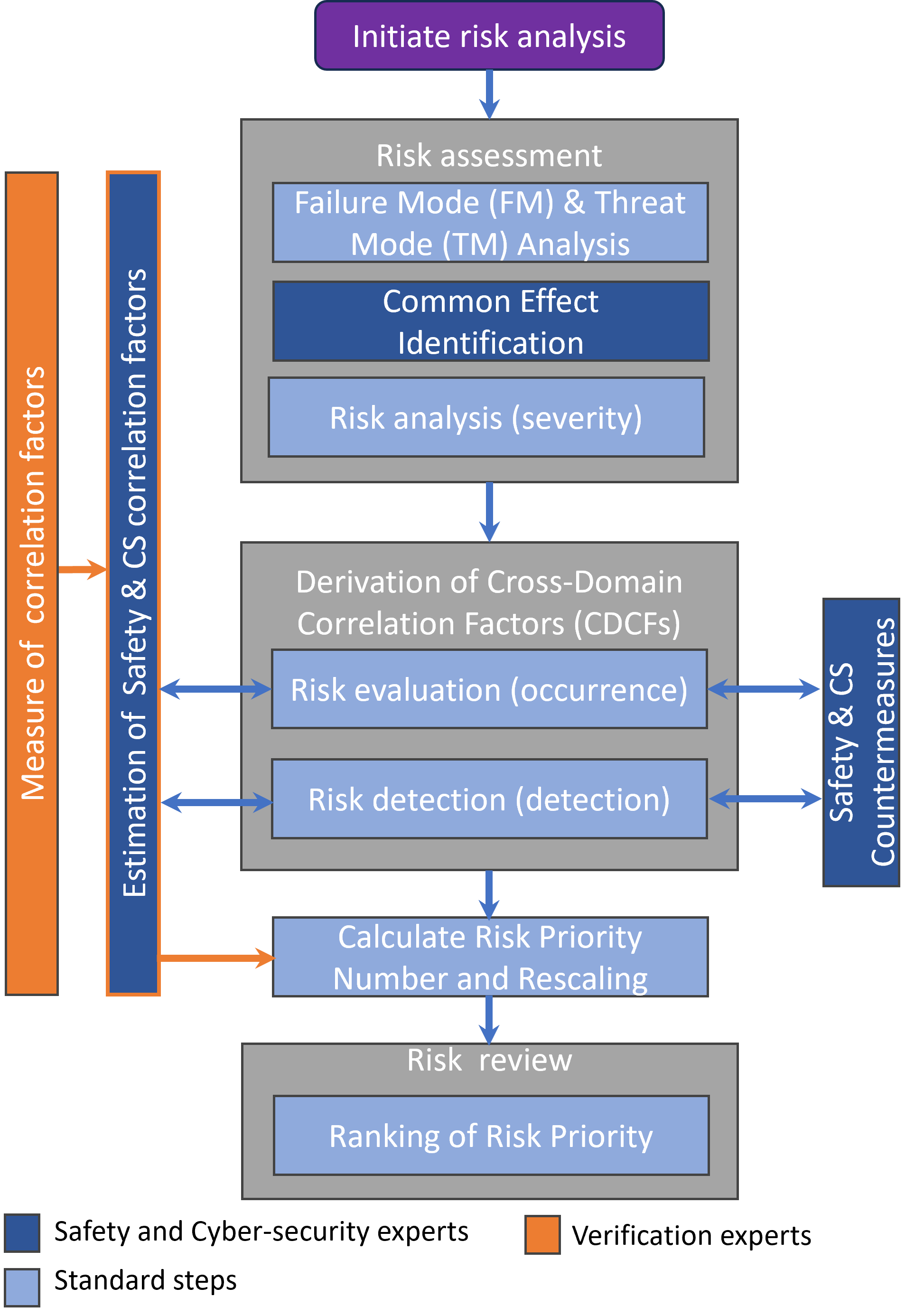}
	\caption{Proposed FTMEA Flow Diagram with Quantified Cross-Domain Correlation Factors}
	\label{fig1}
\end{figure}

To illustrate the proposed concept, a simple flow diagram is presented in Fig. \ref{fig1}.
The FTMEA process is structured as follows (Fig. \ref{fig1}):
\begin{itemize}
\item Initiate Risk Analysis: Define the system, its boundaries, and the critical functions subject to both safety and security requirements. This includes identifying key assets and potential impact scenarios.
\item Failure Mode (FM) and Threat Mode (TM) Analysis: Identify all relevant FMs (e.g., hardware faults, software bugs) and TMs (e.g., cyberattacks, data breaches) for the system or component under analysis. Initial baseline, Occurrence (O) and Detection (D) rates are established. Severity (S) is assessed with the understanding that cybersecurity threats can impact  safety.
\item Common Effect Identification: For each identified FM and TM, analyze their potential adverse effects. A critical step is to identify "Common Effects"—adverse outcomes that can result from either a functional failure or a cybersecurity threat. This forms the basis for cross-domain interaction \cite{b20}.
\item  Derivation of Cross-Domain Correlation Factors (CDCFs): This is the core novel contribution of our framework. CDCFs quantitatively assess the interdependencies between FMs, TMs, and their respective countermeasures ( in dark blue in Fig. \ref{fig1}). This derivation involves a systematic process detailed in Section 3.2.
\item  Calculate Risk Priority Number and Rescaling: The traditional RPN is enhanced by incorporating the derived CDCFs into the Occurrence (O) and Detection (D) ratings. This yields a mathematically robust and more representative RPN for integrated risks, detailed in Section 3.3. Importantly, a systematic rescaling procedure is then applied to map these continuous calculated values back to the discrete 1-10 FMEA ordinal scale.
\item  Ranking of Risk Priority: Prioritize the identified risks based on the modified RPN. Based on this prioritization, develop and implement integrated mitigation strategies that effectively address both safety and security concerns, potentially leveraging cross-domain countermeasures.
\end{itemize}

\subsection{Quantification of Cross-Domain Correlation Factors (CDCFs)}
\begin{figure}[htbp]
	\centering
	\includegraphics[scale=0.5]{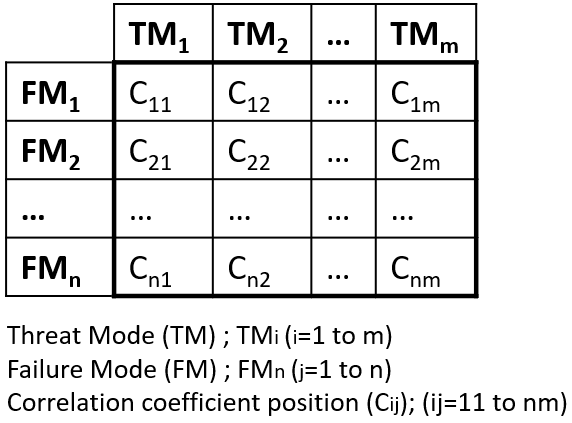}
	\caption{Correlation matrix of common effect where each value is the Cross-Domain Correlation Factor (CDCF)}
	\label{matrice1a}
\end{figure}

A correlation matrix is a table showing correlation coefficients between failure modes and threats as 
presented in Fig. \ref{matrice1a}.
In this way, it is possible to summarize and attempt to quantify the common effects 
between safety and cybersecurity. Each cell value CDCF can be assumed to range between 0 (no correlation) 
and 1 (fully correlation).

Preventive and detection actions are also considered from the cybersecurity point of view to evaluate if such measures can support 
safety. In such case, the safety prevention and detection measure can impact in positive or negative way on 
security and vice versa. 
This phenomenon can be modeled using an asymmetric correlation matrix that 
holistically considers all measures.
For example, a preventive action of security can generate a positive impact on many detection of a safety 
mechanisms. This is the reason that all measures in the correlation matrix shall be evaluated 
together. Each cell $C_{ij}$ (CDCF) of the matrix can be assumed to have a value between 1 (positive impact) 
to -1 (negative impact). From methodological point of view, this kind of analysis can be performed at 
different abstraction levels. For example, at top level of design, considering the top-level measures 
or per parts/subparts \cite{b1} to evaluate local effects.

\subsection{Calculation of RPN}
The traditional Risk Priority Number (RPN) is given by RPN=O×S×D, where O is Occurrence, S is Severity, and D is Detection. All values used to calculate the RPN are set in the range 1-10  \cite{b11}.
The new RPN , reported in the equation \eqref{eq3},  is calculated based on the current state of the art 
\cite{b11} with the difference that the single values of Occurrence and Detection are calculated 
considering the positive and negative impact $C_{ij}$  (see Section 3.2)  according to the equations \eqref{eq1} and \eqref{eq2} 
in relation to the kind of measure, where n are the number of measures that can influence 
the detection or prevention countermeasures. 
\begin{itemize}
	\item Preventive measures: Influence the preliminary evaluation of occurrence of failure/threat 
	mode;
	\item Detection measures: Influence the preliminary evaluation of detection of failure/threat mode;
\end{itemize}

The limits of 1 and 10 in the equations \eqref{eq1} and \eqref{eq2}  are defined to  avoid to deviate from the RPN state of art that can occur when a measure has cumulative effects to  the other domain (eg. preventive action of security impacts in positive way on many safety mechanisms).

\begin{equation}
	\text{$O_{corr}$} =
	\begin{cases} 
		\text{10} & \text{if } \text{$O_{corr}$} \geq 10 \\
		floor(O-O\sum _{j=1}^{n} {C}_{ij}) & \text{if } 1 \leq \text{$O_{corr}$} < 10 \\
		\text{1} & \text{if } \text{$O_{corr}$} < 1
	\end{cases}
	\label{eq1}
\end{equation}

\begin{equation}
	\text{$D_{corr}$} =
	\begin{cases} 
		\text{10} & \text{if } \text{$D_{corr}$} \geq 10 \\
		floor(D-D\sum _{j=1}^{n} {C}_{ij}) & \text{if } 1 \leq \text{$D_{corr}$} < 10 \\
		\text{1} & \text{if } \text{$D_{corr}$} < 1
	\end{cases}
	\label{eq2}
\end{equation}

\begin{equation}
	\text{$RPN$} =\text{$S$}\text{$O_{corr}$}\text{$D_{corr}$} 
	\label{eq3}
\end{equation}

\begin{figure}[htbp]
	\centering
	\includegraphics[width=\linewidth]{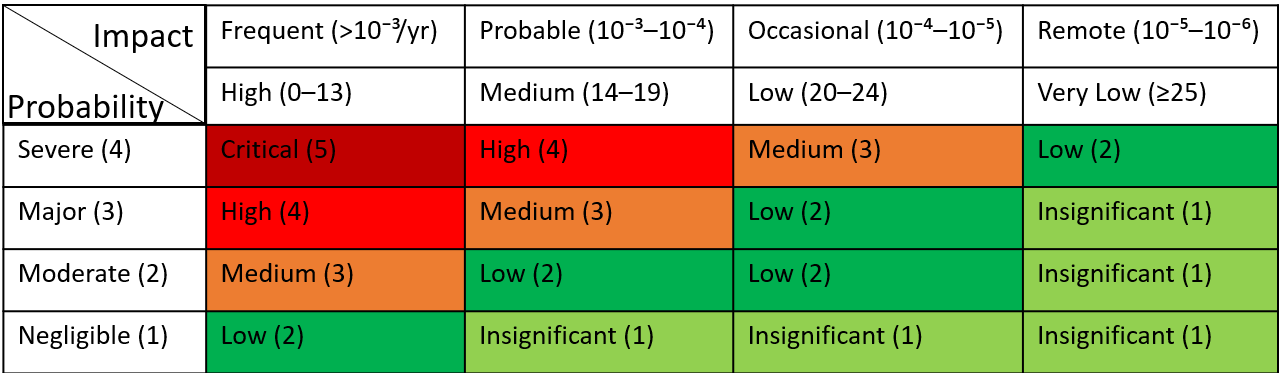}
	\caption{Proposed mapping of risk matrix}
	\label{riskmatrix}
\end{figure}

The probability of occurrence can be derived from the combination of probabilitiy of 
failure and attack probability. Risk can be assessed by considering safety impacts from a 
cybersecurity perspective. Probability of failure (based on ISO 26262 part.11) can be mapped to attack 
potential or 
attack feasibility rating (from ISO 21434), and the risk impact can be evaluated accordingly. Risk 
values can be calculated based a unified risk matrix. The mapping between the proabbility rating and 
safety impact is presented in Fig. \ref{riskmatrix}.

\subsection{Measure of Cross-Domain Correlation Factors (CDCFs)}
Assessing the common effect and mutual influence between prevention and detection measures of safety and CS worlds generally includes observing the IC design by simulation:

\begin{itemize}
	\item Functional Safety: Propagation of the injected faults can be performed to study the effect and the effectiveness of countermeasures;
	\item Cybersecurity: Vulnerability or Penetration tests can be simulated to study the resilience 
	of circuits;
\end{itemize}

The combination of these teqhniques can be used to understand the correlations. Nevetheless, to reduce the effort and the time consuming simulations, structural analysis \cite{b31} and/or Sandia Controllability/Observability Analysis Program (SCOAP) technique \cite{b30} can be used. For example:
\begin{itemize}
	\item Common effect: A super Cone of Influence (COI) can be considered starting from the outputs 
	where the Failure and Threat Effects appear. The difference between the safety and security is 
	that the logic controllables from a cyber attack could not be fully coincident with the random 
	faults that can affect the outputs of COI (Fig.\ref{fig2}). The degree of overalapping between the 
	logic inside the COI and the controllable logic from an attack determines the correlation 
	coefficient.
	
	\begin{figure}[htbp]
		\centering
		\includegraphics[scale=.9]{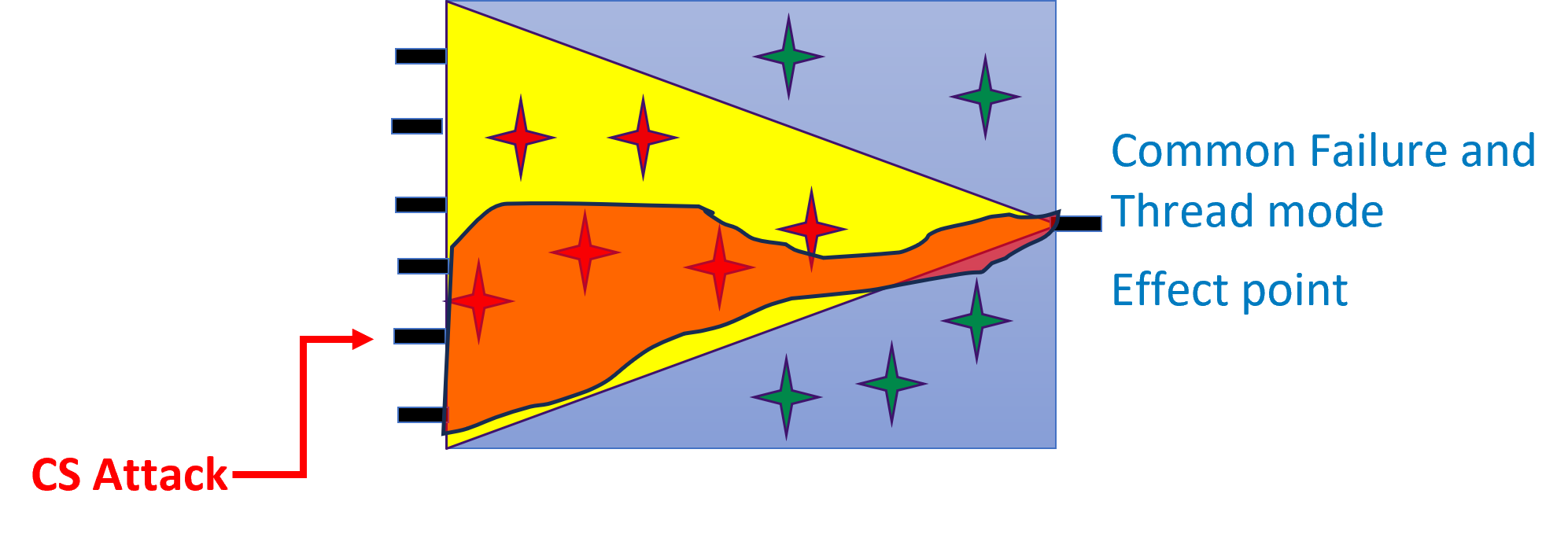}
		\caption{Illustration of the application of structural analysis for deriving CDCF. The "Common Failure and Threat Mode Effect point" highlights where is expected the effect of malfunction. The red highlighted region represents the Cone of Influence (COI) that is common to both a fault propagation path (FM) and an attack vector (CS Attack). The density and criticality of shared logic within this COI, quantifiable via metrics like gate count, provides the basis for the "COI Density Score" contributing to CDCF. A larger and more critical shared region implies a stronger correlation.}
		\label{fig2}
	\end{figure}

	\item Prevention influence: SCOAP technique can be used to calculate the controllability of a 
	circuit respect of inputs used for an attack and the corresponding prevention countermeasure. The 
	controllability can determine the attack resistance similar to  random-pattern resistant 
	(RP-resistant) \cite{b32}. At the same time, the logic involved from the prevetion technique could 
	be rated respect to safety techniques. The area interested from the security prevetion measure can 
	be measured by controllability difference respect to the circuit without measure. For example, a 
	lock and unlock mechanism based on a secret key to prevent write operations on a register change 
	the overall controllability from the input of attack up to the threat effect point 
	(Fig.\ref{fig3}).

	\begin{figure}[htbp]
		\centering
		\includegraphics[scale=0.50]{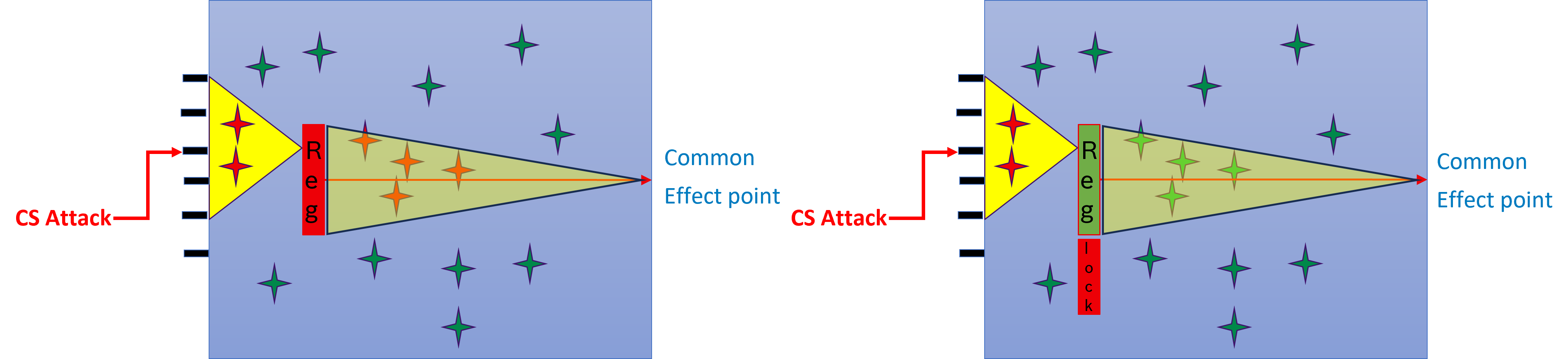}
		\caption{Example how a cybersecurity prevention measure, the register lock mechanism, can affect functional safety by altering hardware controllability. By analyzing the change in the Controllability score   of safety-critical elements, using SCOAP  from the output to the inputs, we quantify the CDCF between the security coutermeasure and safety aspects. In this case,  a security CM that reduces controllability for malicious access and accidental fault propagation will generate a positive CDCF.}
		\label{fig3}
	\end{figure}

	\item Detection influence: In similar way of common effect evaluation, a structural analysis can be done respect to the detection points (Fig.\ref{fig4}). Additionaly, the alarm of safety mechanism from structural point of view could observe part of the security logic with the possibility to analyse negative impact of influence on safety measure.
	\end{itemize}

\begin{figure}[htbp]
		\centering
	\includegraphics[scale=1]{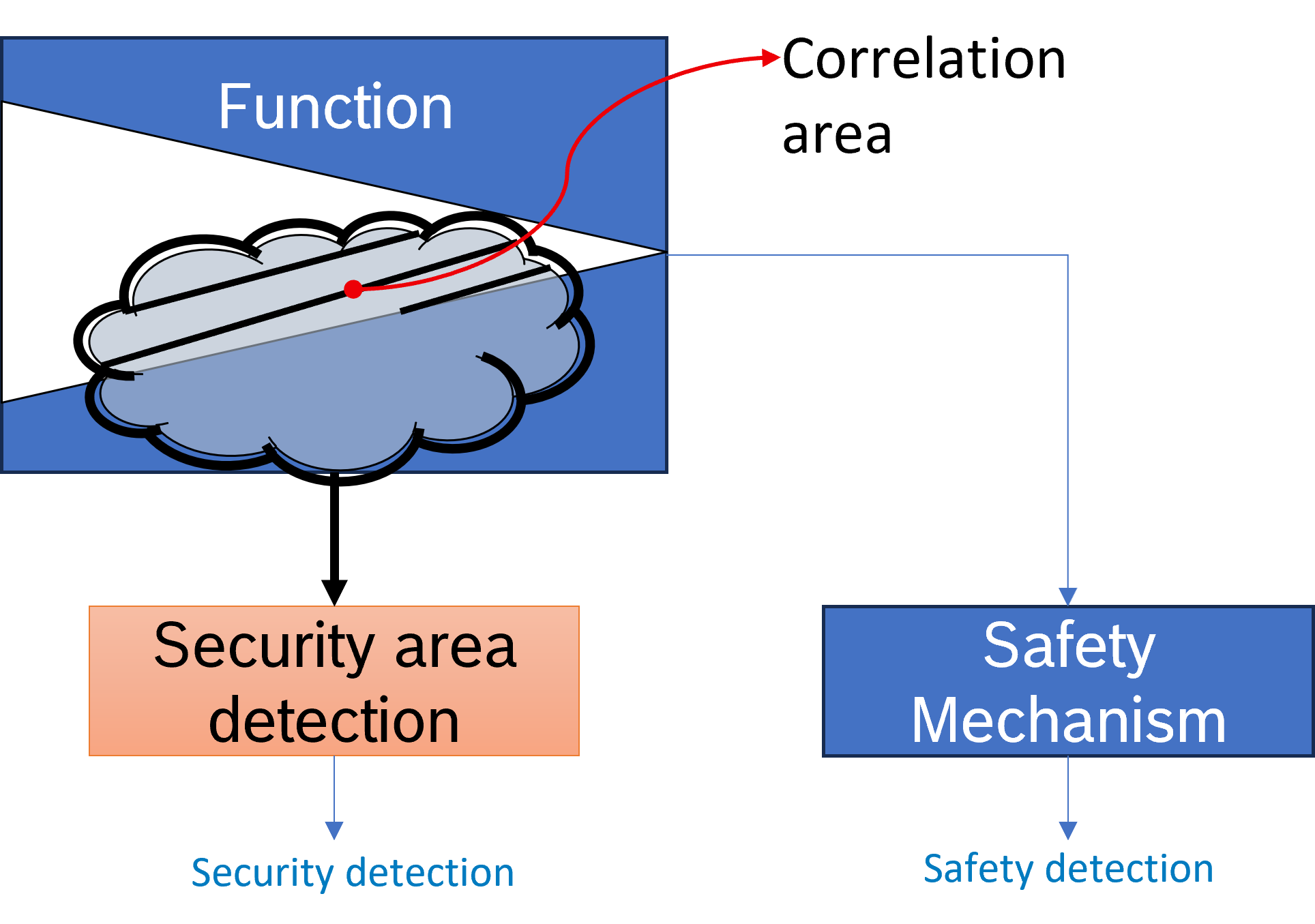}
	\caption{Example how the "Security area detection" and "Safety Mechanism" can share observation points for a "Common Effect point." The "Correlation area" represents the overlap in their monitoring capabilities. Quantification of this overlap (e.g., shared observable registers, common diagnostic paths) via structural analysis contributes to CDCF. A high degree of overlap indicates that a detection mechanism in one domain (e.g., integrity check) can also effectively contribute to detecting issues originating from the other domain, thereby quantifying a positive detection correlation.}
	\label{fig4}
\end{figure}

\subsection{Case Study: Automotive ASIC Configuration Register}
\begin{figure}[htbp]
	\includegraphics[width=\linewidth]{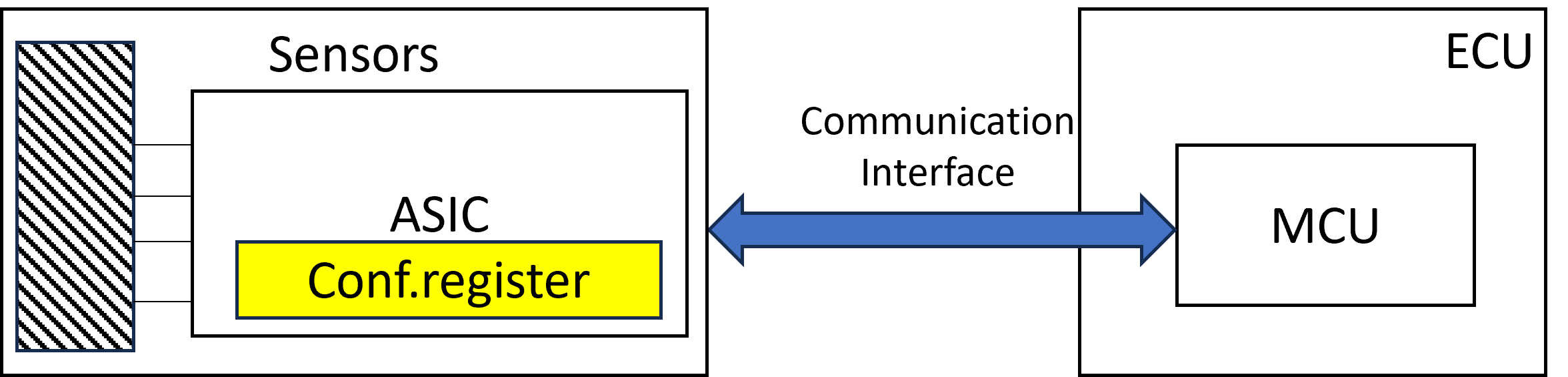}
	\caption{Use Case-Block Diagram. The configuration register, in yellow, is the scope of the analysis.}
	\label{fig5}
\end{figure}

A critical Application Specific Integrated Circuit (ASIC) designed for an automotive sensor module is analyzed by our proposed FTMEA. This ASIC is responsible for transmitting sensor data to the Electronic Control Unit (ECU) (Fig. \ref{fig5}). A key component within this ASIC is a dedicated configuration register (Conf.register). This register stores crucial calibration parameters (e.g., gain, offset, filtering coefficients) that directly affect the accuracy and integrity of the sensor's output. Any unauthorized or erroneous modification to this Conf.register can lead to incorrect sensor data, which can have immediate and severe safety implications (e.g., leading to erroneous brake activation or inaccurate object detection in ADAS).

\begin{figure}[htbp]
	\centering
	\includegraphics[scale=0.8]{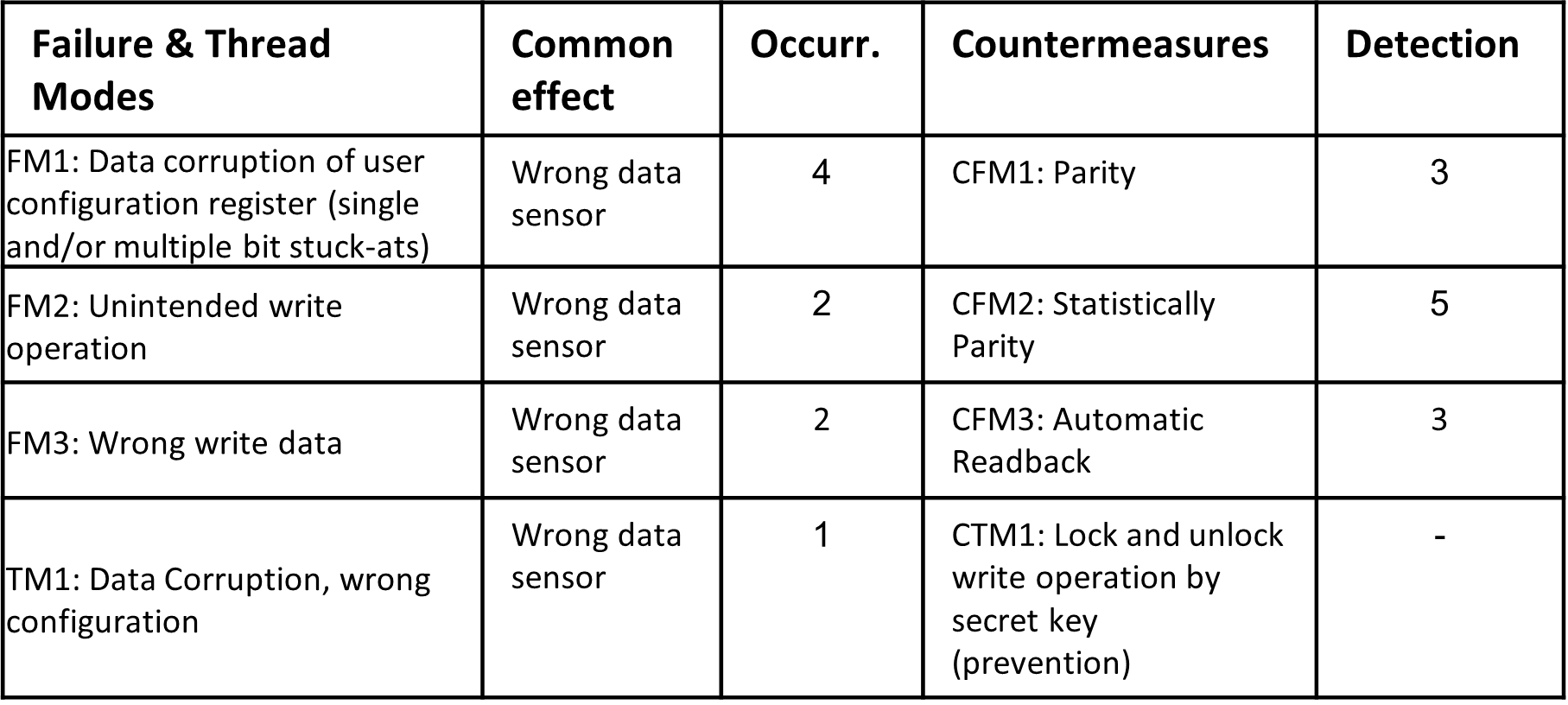}
	\caption{Identified Failure Modes and Threat Modes with Common Effect. The table is a reduced view of the FTMEA related to the configuration register subpart to highlight the common effect and the used measures.}
	\label{failure_thread}
\end{figure}

Focusing the analysis on the configuration register, the Fig.\ref{failure_thread} reports the 
failure and threat modes that produces the same overall effect on the sensor.

\begin{figure}[htbp]
	\centering
	\includegraphics[scale=0.65]{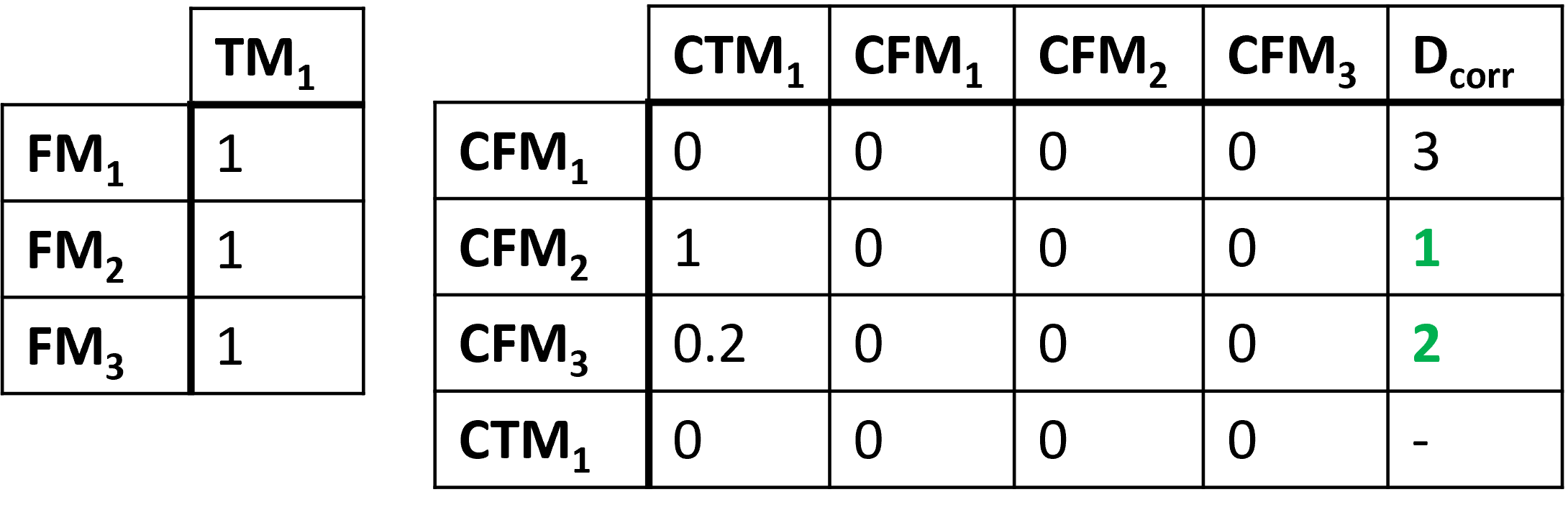}
	\caption{Calculation of Correlation matrix of common effect based on the use case. CTM=CDCF Threat Mode, CFM=CDCF Failure Mode. The  \text{$D_{corr}$} is calculated by  the equation \eqref{eq2}  where $C_{ij}$ are the elements of the matrix.}
	\label{detection_measures}
\end{figure}

All the proposed countermeasures are related to detection, whereas the security 
measure functions as a prevention mechanism and for this reason  the Occurrence is 1.
The  detection correlation matrix can be done as illustrated in
Fig.\ref{detection_measures}.  In detail, 
the parity can statistically detect an "Unintend write operation"
if it occurs with multiple bit erros. The usage of lock and unlock mechanisms fully detect this case, 
so the measure for the FM2 is fully covered because change the controllability of the register. Even 
in the case of FM3 the threat countermeasure can partially detect a problem in the logic devoted to handle the write operations on the registers. 
From a measure of observability respect to the threat countermeasure alarm, the write interface is 
visible. All measures have been performed on the gate-level netlist using a synthesis tool and 
proprietary scripts to prove the CDCF factors. The  Detection of in Fig \ref{detection_measures} is calculated using 
the  equations  \eqref{eq2}.

\begin{figure}[htbp]
	\centering
	\includegraphics[scale=0.75]{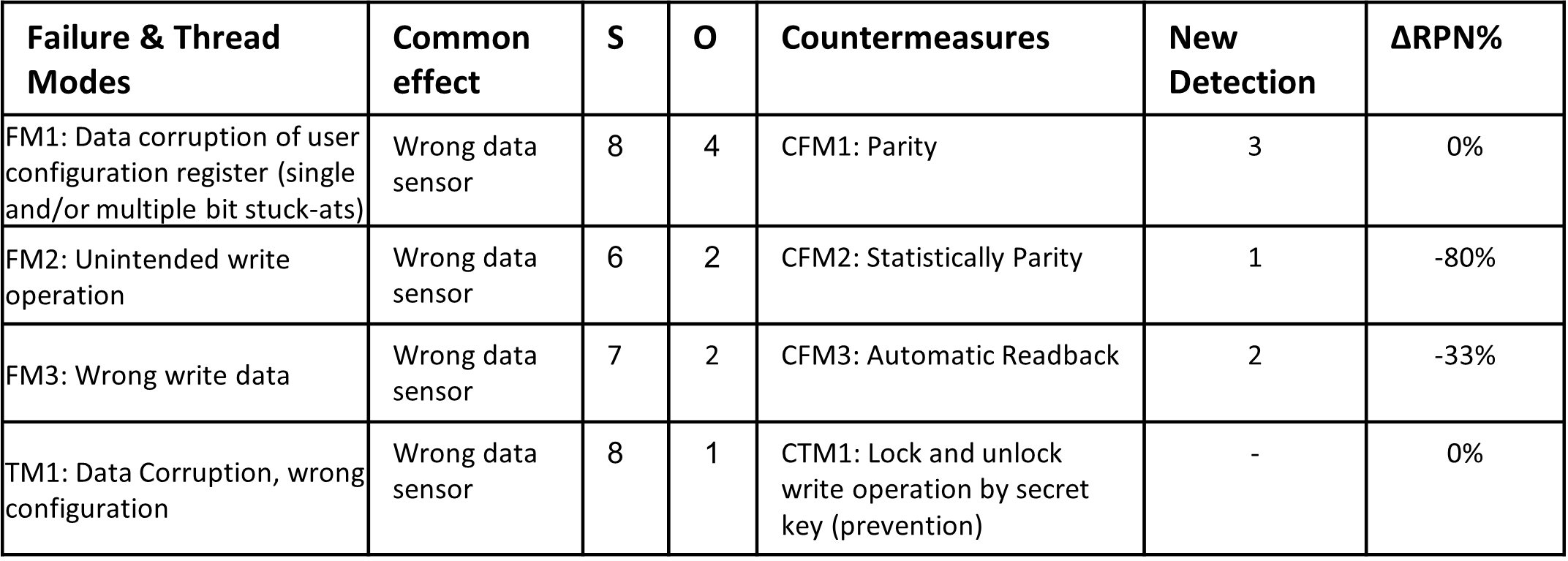}
	\caption{Recalculation of RPN   based on \eqref{eq3} by the  \text{$D_{corr}$} in the Correlation matrix.}
	\label{RPN_recalculation}
\end{figure}

The Fig \ref{RPN_recalculation} demonstrates the impact and value of the FTMEA framework.  The RPN  calculation, based on \eqref{eq3},  uses the modified Detection ($D_{corr}$) values, incorporating the derived CDCFs from Fig \ref{detection_measures}, and reflecting the integrated effect of both FuSa and Cybersecurity countermeasures.  The last column of Fig \ref{RPN_recalculation} highlights the improvement in percentage of the RPN. Such improvements allow to the team to reduce effort and the integration of additional measures in the design.

\section{Discussion of Results}
The comparative analysis in Fig \ref{RPN_recalculation} clearly demonstrates the value and impact of the FTMEA framework:
\begin{itemize}
	\item Revealing Overestimated Risks: Failures FM2 and FM3.
           \item Quantifying Synergistic Countermeasure Benefits: The correction of Detection has created benefit on  FM2 and FM3.
           \item Improved Prioritization for Resources:  The failures that are not affected from synergistic benefit can be properly attentioned for evaluate additional measures.
           \item  CDCFs Methodology: The systematic shifts in RPN, which are directly traceable and attributable to the specific CDCF values derived from structural analysis,  or simulation, provide strong quantitative evidence for the validity and operational utility of our proposed methodology. The changes are logical, consistent, and explainable by the underlying correlations, thus validating the framework's ability to model real-world interactions. 
\end{itemize}
In summary, the FTMEA framework with its quantified CDCFs provides a more comprehensive, accurate, and actionable risk landscape. It enables design teams to identify and prioritize truly integrated risks, moving beyond qualitative assessments to data-driven decisions for optimal resource allocation in designing resilient and secure automotive systems.

\section{Conclusion}
This paper introduced a novel Integrated Failure and Threat Mitigation Analysis (FTMEA) framework that systematically addresses the intertwined risks of functional safety and cybersecurity in complex automotive semiconductor devices. Our core contribution lies in the rigorous definition, systematic derivation, and empirical quantification of Cross-Domain Correlation Factors (CDCFs). These CDCFs model the mutual influences between functional failure modes, cybersecurity threat modes, and their respective countermeasures, moving beyond qualitative acknowledgments to verifiable numerical values. By integrating these CDCFs into a mathematical modification of the Risk Priority Number (RPN) calculation, followed by a transparent rescaling procedure, we have developed a reproducible, traceable, and empirically justifiable approach to holistic risk assessment 
The detailed case study, involving an automotive ASIC configuration register, demonstrated the practical application of FTMEA with explicit data and CDCF derivations.
This unified analysis framework promotes cross-disciplinary collaboration and facilitates the development of more 
resilient semiconductor designs. As a result, it enhances both compliance with safety standards (e.g., 
ISO 26262) and alignment with cybersecurity guidelines (e.g., ISO/SAE 21434).
This holistic approach helps developers trace complex 
information and failure flows across both top-level and bottom-level phases during system 
development. It also supports distributed development across different disciplines and organizations 
by providing a structured way to manage interdependencies and ensure consistency between functional 
safety and cybersecurity requirements.
While the proposed FTMEA framework offers significant advancements, certain limitations must be acknowledged: Deriving highly accurate CDCFs requires substantial investment in expert elicitation, sophisticated structural analysis tools, and dedicated fault/attack injection campaigns, which can be resource-intensive, particularly for early design phases.
A future work will be devoted to apply the proposed approach on a complex use case to improve and compare the measure of coefficients, in this work based on structural circuit analysis, with dynamic simulations to 
evaluate the potential gaps. Investigating the application of machine learning and artificial intelligence techniques to partially automate the correlation factor derivation  will be also investigated.

%
%







\begin{thebibliography}{00}
	\bibitem{b1} ISO 26262: Road vehicles – Functional safety, Part 11: Guidelines 
	on appliction of 
	ISO 26262 to semiconductors, 2018
	
	\bibitem{b2} ISO 21434: Road vehicles – Cybersecurity engineering, 2021
	
	\bibitem{b3} S. M. Nicoletti, M. Peppelman, C. Kolb and M. Stoelinga, 
	"Model-based Joint Analysis of 
	Safety and Security: Survey and Identification of Gaps," Computer Science 
	Review, vol.50, issn 
	1574-0137, 2023
	
	\bibitem{b4} A Abdulhamid, S. Kabir, I. Ghafir and C. Lei, "An Overview of 
	Safety and Security 
	Analysis Frameworks for the Internet of Things," Electronics,vol.12, 2023
	
	\bibitem{b5} SAE J3061: Cybersecurity Guidebook for cyber-Physical 
	Vehicle Systems, 2021
	
	\bibitem{b6} ISO/TR 4804: Road vehicles – Safety and cybersecurity for 
	automated driving systems – 
	Design, verification and validation, 2020
	
	\bibitem{b7} ISO TR 63069: Industrial-process measurement, control and 
	automation-Framework for  
	functional safety and security, 2019
	
	\bibitem{b8} ISO 21448: Road vehicles — Safety of the intended functionality, 2022
	
	\bibitem{b9} ISO 24089: Road vehicles — Software update engineering
	
	\bibitem{b10} E. Rama et al., "Trustworthy Integrated Circuits: From Safety to 
	Security and Beyond," 
	in IEEE Access, vol. 12, pp. 69603-69632, 2024, doi: 
	10.1109/ACCESS.2024.3400685.
	
	\bibitem{b11} N. Sellappan and K. Palanikumar, "Modified Prioritization Methodology for Risk Priority 
	Number in Failure Mode andEffects Analysis," 
	in International Journal of Applied Science and Technology, vol. 3 , no. 4, 2013.
	
	\bibitem{b12} C. Schmittner, T. Gruber, P. Puschner, E. Schoitsch, "Security 
	Application of Failure 
	Mode and Effect Analysis (FMEA)," A. Bondavalli and F. Di Giandomenico (eds.), 
	Computer 
	Safety, Reliability, and Security, Springer International Publishing, Lecture 
	Notes in Computer 
	Science, vol 8666. Springer, Cham., 2014, doi: 10.1109/ISSE63315.2024.10741112.
	
	\bibitem{b13} V S Bharath Kurukuru and Irfan Khan, "Failure Mode Effect 
	Analysis of Power 
	Semiconductors in a Grid‐Connected Converter," in Fault Analysis and its Impact 
	on Grid-connected 
	Photovoltaic Systems Performance , IEEE, 2023, pp.149-184, doi: 
	10.1002/9781119873785.ch5.
	
	\bibitem{b14} C. Binder, S. Hoher, B. Maxim, S. Riedmann, C. Neureiter and S. 
	Huber, "Application of 
	FMVEA by Design for Adding Smart Functionalities to an Existing Campervan," 
	2024 IEEE International 
	Symposium on Systems Engineering (ISSE), Perugia, Italy, 2024, pp. 1-6, doi: 
	10.1109/ISSE63315.2024.10741112.
	
	\bibitem{b15} R. Chemali, B. Conrard, M. Bayart. FVMEARA : A NEW SYSTEMATIC 
	APPROACH for SECURITY and 
	SAFETY RISK Co-ASSESMENT BASED ON ICVSS METHODOLOGY. 13ème CONFERENCE 
	INTERNATIONALE DE 
	MODELISATION, OPTIMISATION ET SIMULATION (MOSIM2020), 12-14 Nov 2020, Morocco. ⟨hal-03190661⟩
	
	\bibitem{b16} Y. Chen, X. Q. He and P. Lai, "The application of fault tree 
	analysis method in electrical component," Proceedings of the 20th IEEE 
	International Symposium on the Physical and Failure Analysis of Integrated 
	Circuits (IPFA), Suzhou, China, 2013, pp. 658-661, doi: 
	10.1109/IPFA.2013.6599246.
	
	\bibitem{b17} I. Friedberg, K. MxLaughlin, P. Smith, D. Laverty and S. Sezer, 
	"STPA-SafeSec: Safety and security analysis for cyber-physical systems," 
	Journal of Information Security and Applications, Vol. 34, PP-183-196, 2017
	
	\bibitem{b18} N. Ali,  M. Hussain, and J.-E. Hong, "SafeSoCPS: A Composite 
	Safety Analysis Approach for System of Cyber-Physical Systems," journal Sensors,
	Vol. 22, Nr.12, ArticleNr. 4474, 2022, doi:10.3390/s22124474
	
	\bibitem{b19} G. Sabaliauskaite, L.S. Liew, and F. Zhou," AVES – Automated 
	Vehicle Safety and Security Analysis Framework," Proceedings of the 3rd ACM 
	Computer Science in Cars Symposium, Association for Computing Machinery, New 
	York, NY, United States, 2019, doi: 10.1145/3359999.3360494
	
	
	\bibitem{b20} M. Khatun, A. Armato, and S. Fischer, "A Holistic Approach to 
	Investigate Failure Effects
	for Analysing Functional Safety and Cybersecurity
	in Semiconductors," 8th International Conference on System 
	Reliability and Safety (ICSRS 2024), Sicily, Italy, November, 2024, pp. 623-628, doi: 
	10.1109/ICSRS63046.2024.10927504.
	
	\bibitem{b21} AIAG \& VDA FMEA-Handbook, Failue mode and Effect analysis-FMEA Handbook, Design FMEA 
	and process FMEA, Supplimental FMEA for Monitoring \& system Response,Germany, 2022 
	
	\bibitem{b22} J. Hartwell, "FMEA RPN – Risk Priority Number. How to Calculate and Evaluate," FMEA 
	Studio, 2022
	
	\bibitem{b23} UN Regulation No. 156 - Software update and software update management system,
	E/ECE/TRANS/505/Rev.3/Add.155, March,2021 
	
	\bibitem{b24} ANSI/UL 4600: Standard for Evaluation of Autonomous Products, edi.3, March, 2023.
	
	\bibitem{b25} UN Regulation No. 155 - Cyber security and cyber security management system, 
	E/ECE/TRANS/505/Rev.3/Add.154, March,2021
	
	\bibitem{b28}	M. Khatun, F. Wagner, R. Jung and M. Glass, "An Approach 
	of a Safety Management System for Highly Automated Driving System," 
	2021 5th International Conference on System Reliability and Safety 
	(ICSRS), Palermo, Italy, 2021, pp. 222-229, doi: 
	10.1109/ICSRS53853.2021.9660687.
	
	\bibitem{b29} Sun, Yuzhi, et al. "Dynamic failure modes and effects analysis method considering synergistic standards of failure correlation and expert evaluation." Expert Systems with Applications 279 (2025): 127461.
	
	\bibitem{b30} Goldstein, Lawrence H., and Evelyn L. Thigpen. "SCOAP: Sandia controllability/observability analysis program." Proceedings of the 17th Design Automation Conference. 1980.
	
	\bibitem{b31} da Silva, F. Augusto, et al. "Use of Formal Methods for verification and optimization of Fault Lists in the scope of ISO26262." 2018 Design and Verification Conference and Exhibition (DVCon) Europe, Munich, Germany. 2018
	
	\bibitem{b32} J. Savir, G. S. Ditlow, and P. H. Bardell, Random pattern testability, IEEE Trans.Comput., C-33(1), 79–90, 1984
          
           \bibitem{b33} ISO/TS 5083:2025 Road vehicles — Safety for automated driving systems — Design, verification and validation
	







\end{thebibliography}
\end{document}